\documentstyle[amssymb,psfig,prd,aps]{revtex}             \textheight 9.45in

\newcommand\beq{\begin{equation}}
\newcommand\eeq{\end{equation}}
\newcommand\beqa{\begin{eqnarray}}
\newcommand\eeqa{\end{eqnarray}}

\newcommand\sgn{\text{sgn}}
\newcommand\ox{\otimes}

\newcommand{{\lbb}}{[\hskip -1.5pt [}		
\newcommand{{\rbb}}{]\hskip -1.5pt ]}		

\newcommand\rbar{\bar r}
\newcommand\Ubar{\bar U}
\newcommand\rhobar{\bar \rho}

\newcommand\rtil{\tilde r}
\newcommand\ttil{\tilde t}
\newcommand\Util{\tilde U}
\newcommand\rhotil{\tilde \rho}

\newcommand\Gcheck{\check G}
\newcommand\mcheck{\check m}
\newcommand\rcheck{\check r}
\newcommand\tcheck{\check t}
\newcommand\Tcheck{\check T}
\newcommand\Ucheck{\check U}
\newcommand\Vcheck{\check V}
\newcommand\Xcheck{\check X}
\newcommand\rhocheck{\check \rho}

\newcommand\Ehat{\hat E}
\newcommand\Ghat{\hat G}
\newcommand\Rhat{\hat R}
\newcommand\That{\hat T}
\newcommand\Vhat{\hat V}
\newcommand\omegahat{\hat \omega}
\newcommand\Phihat{\hat \Phi}
\newcommand\Psihat{\hat \Psi}

\newcommand\Ao{\mathaccent"17 A}
\newcommand\Fo{\mathaccent"17 F}
\newcommand\go{\mathaccent"17 g}
\newcommand\Go{\mathaccent"17 G}
\newcommand\phio{\mathaccent"17 \phi}
\newcommand\Psio{\mathaccent"17 \Psi}

\newcommand\degree{\mathaccent"17 {}}

\input epsf

\begin{document}

\draft

\wideabs
{ {
\title
{
Darkholes: Nicer than blackholes --- with a bright side, too\footnotemark \\
(Does energy produce gravity?)
}

\author{Homer G. Ellis}
\address{Department of Mathematics, University of Colorado at Boulder,
Boulder, Colorado 80309}
\date{February 23, 2000}
\maketitle

\begin{abstract}
The geometry of three-dimensional space guides the search for a better model
than the blackhole with its unwelcome singularity.  An elementary construction
produces on the 4-manifold of 2-spheres in a Riemannian 3-space a space-time
metric invariant under uniform conformal transformations of the 3-space.  When
the 3-space is Euclidean, the metric reduces to de Sitter's expanding universe
metric.  Generalization yields a space-time metric that retains the
`exponential expansion property' of the de~Sitter metric.  A strictly geometric
action principle gives field equations which, because they do not adhere to
Einstein's early confounding of energy and inertial mass with gravitating mass,
admit solutions that escape the Penrose--Hawking singularity theorems.  A
spherically symmetric solution that is asymptotic to the Schwarzschild
blackhole metric has, in place of a horizon and a singularity, an
Einstein--Rosen `bridge', or `tunnel', connecting two asymptotically Euclidean
regions.  On one side the gravitational center attracts, and is dark but not
black; on the other side it repels, and is bright.  Travel and signaling from
either side to the other {\em via} the tunnel are possible.  Analysis of the
Einstein tensor of this `darkhole' (or `darkhole--brighthole') suggests that
not all energy produces gravity, and that calling energy `negative', or its
relationship to geometry `exotic', is unjustified.
\end{abstract}

} }

\narrowtext

\footnotetext{\footnotemark Revision and
amplification of Darkholes: Blackholes' Better Behaved Cousins, selected for
honorable mention in the Gravity Research Foundation 1999 Awards for Essays
on Gravitation.}

\renewcommand{\thefootnote}{\arabic{footnote}}
\setcounter{footnote}{0}

Blackholes with singularities are not satisfactory models of real things, for
at the singularities they lose their predictive powers, causing one to throw up
the hands and mutter some such incantation as ``Quantum effects take over.''
It is easy to construct a blackhole without a singularity: if, for
$-\infty < \rho < \infty$,
\beq
\Ghat = dt^2 - [d\rho - u(\rho)\, dt]^2 - r^2(\rho)\, d\Omega^2,
\label{flowform1}
\eeq
where $d\Omega^2 := d\vartheta^2 + (\sin \vartheta)^2 d\varphi^2$, then
$\Ghat$ has no singularity, provided that neither $u(\rho)$ nor $r(\rho)$ has,
that $u(\rho)$ and $r(\rho)/\rho$ are bounded at $-\infty$ and at $\infty$, and
that $r(\rho)$ is bounded away from 0.  Blackness occurs if there is a region
in which $|u(\rho)| > 1$, bounded by a sphere (or two spheres) on which
$|u(\rho)| = 1$ ---~for the following reasons:  Photon orbits are characterized
by the equations
\beq
\frac{d\rho}{dt} = u(\rho) \pm \sqrt{1 - r^2(\rho)
                                         \left(\frac{d\Omega}{dt}\right)^2}.
\eeq
In a region where $|u(\rho)| > 1$, either $u(\rho) > 1$ throughout or
$u(\rho) < -1$ throughout.  In the first case $d\rho/dt > 0$, in the second,
$d\rho/dt < 0$.  In either case all photons in that region are going in the
same direction radially.  None can have entered the region through a bounding
sphere that all are approaching, and none can leave it through a bounding
sphere from which all are retreating.  In each case the sphere is a horizon
for light, thus also for test particles moving slower than light.

The coordinate transformation $T = t + \int u(\rho)\, [1 - u^2(\rho)]^{-1}\,
d\rho$ changes the expression of $\Ghat$ to
\vskip 5pt
\beqa
\lefteqn
{
 \Ghat = \left[1 - u^2(\rho)\right] dT^2
} \qquad \nonumber \\ 
  & & {} - \left[1 - u^2(\rho)\right]^{-1} d\rho^2
                               -  r^2(\rho)\, d\Omega^2.
\label{schwarzform}
\eeqa
\vskip 5pt
\noindent If $r(\rho) \sim \rho$ and $u^2(\rho) \sim 2m/\rho$ as
$\rho \to \infty$, the metric behaves like the Schwarzschild metric of mass
parameter $m$, so it can model the far field of a spherically symmetric
gravitating object.  Because $r^2(\rho)$ stays away from 0, any region interior
to a horizon is spacious: it does not squeeze down to a point at which a
singularity could develop, as the Schwarzschild inner region does.  A
construction analogous to the Kruskal--Fronsdal extension of the Schwarzschild
metric would show that each horizon serves as a neck of a wormhole connecting
two or more regions in which $|u(\rho)| < 1$.

Such a singularity-free blackhole cannot, of course, be a solution of the
Einstein field equations.  It must in fact escape in some way the
Penrose--Hawking singularity theorems~\cite{haw&ell73}, and this it can do only
by violating one of the hypotheses of those theorems.  Among those hypotheses
suspicion attaches most readily to the requirement that the Ricci tensor be
everywhere nonnegative definite with respect to null or timelike vectors.  This
so-called `energy condition' is conventionally taken to mean that the density
of energy, in whatever form, that `produces' a gravitational field must be
nowhere on balance negative.  As `negative energy' is believed to be an
attribute only of never observed `exotic' matter, the energy condition is
almost universally accepted as realistic.  That acceptance, however, rests
ultimately on a questionable identification that traces all the way back to
Einstein's 1916 paper Die Grundlage der allgemeinen
Relativit\"atstheorie~\cite{ein16}. 

In that paper's \S 16, titled in translation The General Form of the Field
Equations of Gravitation, Einstein seeks a tensorial equation to correspond to
the Poisson equation $\nabla^2 \phi = 4 \pi \kappa \rho$, where $\rho$ denotes
the ``density of matter''.  Drawing on the special theory of relativity's
identification of ``inert mass'' with ``energy, which finds its complete
mathematical expression in . . . the energy-tensor'', he concludes that
``we must introduce a corresponding energy-tensor of matter
$\text{T}^\alpha_\sigma$\,''.  Further describing this energy-tensor as
``corresponding to the density $\rho$ in Poisson's equation'', he goes on to
invent the field equations $E_{\mu \nu} = \kappa T_{\mu \nu}$ that bear his
name and have the built-in consequence that, wherever energy density is
nonnegative for all observers, the Ricci tensor is nonnegative definite with
respect to null and timelike vectors (here $E$ is the Einstein tensor
$\Phi - \frac12 \Psi G$, where $G$ is the metric tensor, $\Phi$ is the Ricci
tensor, and $\Psi$ is the curvature scalar; see Appendix for definitional
conventions).

The questionable identification referred to is the confounding of `gravitating
mass', which is the sole contributor to the ``density of matter'' in Poisson's
equation, with ``inert mass'', thus with energy by way of $E = m c^2$.  That
all bodies respond alike to a gravitational field establishes the equivalence
of `{\em passive}' (gravitat{\em ed}\/) mass with `inertial' (inert) mass, but
an equivalence between `{\em active}' (gravitat{\em ing}\/) mass and
passive(--inertial) mass is in no way implied.  The distinction between
active mass and passive mass, well explicated by Bondi \cite{bon57}, is
present already in Newton's gravitational equation
$m_{\text{inertial}}\, a = -G\, M_{\text{active}}\, m_{\text{passive}}/r^2$,
where $M_{\text{active}}$ and $m_{\text{passive}}$ are properties of entirely
different bodies, one doing the acting, the other receiving the
action.\footnote{That Einstein confounded active mass with passive--inertial
mass, knowingly or unknowingly, is borne out further by the statement in
his~\S 16 that for a ``complete system (e.g.~the solar system), the total mass
of the system, and therefore its total gravitat{\em ing} action as well, will
depend on the total energy of the system, and therefore on the {\em ponderable}
energy together with the gravitational energy.'' (Emphasis added.)}

If active mass is not equivalent to passive mass, it is not equivalent to
inertial mass, thus is not equivalent to energy.  Unresolved, therefore, is
whether all constituents of matter and energy gravitate, and of those that do,
whether they attract or repel gravitationally.  In an experiment by Kreuzer
\cite{kre68}, two congruent, homogeneous bodies, differently constituted but
weighing the same, were seen to exert the same gravitational attraction on test
particles (within experimental precision).  This indicates equality of the
ratio of active to passive mass for the two macroscopic bodies, but it says
nothing about the gravitational effects of energy, or of any particular species
of the particles that make up matter.  It is consistent with this observation
to suppose, for example, that only nucleons produce gravitational effects, that
energy and other particles such as electrons and neutrinos do not gravitate at
all.  To see this, consider an idealized Kreuzer experiment in which body A is
is made of a single isotope of one element, each of whose atoms has
$p_{\text A}$ protons, the same number of electrons, and $n_{\text A}$
neutrons, and body B is made of a single isotope of another element, each atom
of which has $p_{\text B}$ protons and electrons, and $n_{\text B}$ neutrons,
with $p_{\text A\!} + n_{\text A} = p_{\text B} + n_{\text B}$, and
$p_{\text A} > p_{\text B}$.  Next, perform the thought experiment of reversing
beta decay in each atom of body A by stuffing \,$p_{\text A} - p_{\text B}$\,
of its atomic electrons, along with as many antineutrinos, into its nuclear
protons, thus turning the protons into neutrons and the A atoms into B atoms,
maintaining congruence all the while.  Now the bodies are identical, and their
weights are still the same --- but so are their active masses, despite that
antineutrinos have been added and binding energies have changed.  It is
conceivable that the binding energies and the antineutrinos have increased A's
active gravitational mass, and that this increase is exactly compensated by a
decrease owed to a loss of molecular kinetic energy necessary to maintain A's
size and weight.  It is also conceivable that they have {\em de}creased A's
active mass, and that this is compensated by an increase of kinetic energy.
It is, however, equally conceivable (and from a probabilistic standpoint even
more likely) that the binding energies, the antineutrinos, and the kinetic
energy produce {\em no} gravity --- that only the nucleons and perhaps (but
perhaps not) the electrons have nonzero active gravitational mass.  Any
contradiction of this in the form of a measurement of the gravity of an
isolated electron, antineutrino, or quantum of energy would seem a distant
prospect at best.  Absent such a measurement, the `energy condition' is an
unproven hypothesis,~nothing\-~more.

In what rational way might one replace the Einstein field equations with others
that allow violations of the `energy condition'?  Geometry should be the guide,
according to Einstein, who likened his equations to a building with two wings,
one made of fine marble (the geometrical tensor), the other of low-grade wood
(the matter tensor) \cite{ein50}.  All the better, a purist says, if it is the
geometry of real three-dimensional space, not the pseudo-geometry of space-time
in which `time' is a fourth dimension, independent of and unrelated to the
three spatial dimensions.  Precisely that geometry is the guide for the
construction that follows.

\begin{figure}[htb]
\psfig{figure=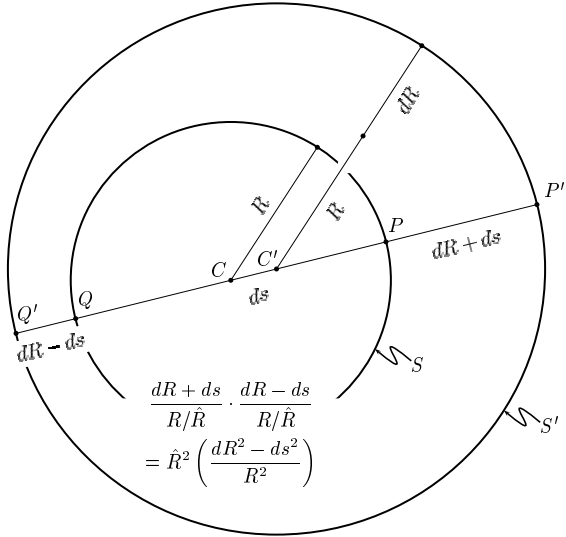,width=3.375truein}
\vskip 10pt

\caption{Neighboring 2-spheres $S$ and $S'$ in the Riemannian 3-space
\{${\cal M}, \Go$\}, shown in cross section through the geodesic $Q'QCC'PP'$,
separated by the `two-point' distance $\Rhat \sqrt{(dR^2 - ds^2)/R^2}$.}
\label{fig1}
\end{figure}

Let $\Go$ be a positive definite Riemannian metric on a three-dimensional
manifold $\cal M$ that is geodesically complete with respect to $\Go$.  The
2-sphere in $\cal M$ of radius $R$ centered at the point $C$ is the set of all
points whose distance from $C$ along a geodesic is $R$.  The set of all such
spheres is itself a four-dimensional manifold $\hat{\cal M}$.  Let $S$ and $S'$
be neighboring spheres in $\cal M$, centered at $C$ and $C'$, of radii $R$ and
$R + dR$.  Starting at $C$ and following a geodesic through $C'$ one arrives at
a point $P$ on $S$ and a point $P'$ on $S'$ separated by a (directed) geodesic
distance $dR + ds$, where $ds$ is the geodesic distance from $C$ to $C'$
(see Fig.~\ref{fig1}).  Going in the other direction one arrives at points $Q$
on $S$ and $Q'$ on $S'$ separated by $dR - ds$.  The product of these
separations, each normalized by division by $R/\Rhat$, where $\Rhat$ is a
positive constant, provides a normal hyperbolic metric $\Ghat$ on
$\hat{\cal M}$ that is invariant under all uniform conformal transformations
of $\Go$ ($\Go \to k \Go$ with $k$ a positive constant), viz.,
\beq
\Ghat = \Rhat^2 \left(dR^2 - ds^2 \over R^2 \right).
\eeq

Assigning to each 2-sphere in $\cal M$ a time $t$ related to its radius $R$ by
$t := -\ln (R/\Rhat)$ gives $\Ghat$ the form
\beq
\Ghat = \Rhat^2\, dt^2 - e^{2t}\, ds^2.
\eeq
Upon particularization of $\Go$ to be the metric of Euclidean 3-space, $\Ghat$
reduces to the metric of de Sitter's expanding universe model \cite{sch56},
a solution of the Einstein vacuum equation
\beq
\Ehat := \Phihat - \case{1}{2} \Psihat \Ghat = \Lambda \Ghat
\eeq
with cosmological constant $\Lambda = 3/\Rhat^2$; $\Rhat$ is the
uniform space-time radius of curvature of this empty universe.

In tensor product form
\begin{mathletters}
\beqa
\Ghat &=& \Rhat^2 (dt \ox dt) - e^{2t}\, \Go  \\
      &=& \Rhat^2 (dt \ox dt) - e^{2t} (dx^m \ox \go_{mn}\, dx^n).
\eeqa
\end{mathletters}
There is on $\hat{\cal M}$ a vector field $\xi$, namely,
$\xi := \partial / \partial t$, with respect to which $\Ghat$ has the following
`exponential expansion property': ${\cal L}_{\xi} \Ghat = 2\, G$, where
$G := \Ghat - (\Ghat \xi \xi)^{-1}(\Ghat \xi \ox \Ghat \xi)$,
${\cal L}_\xi$~denoting Lie differentiation along $\xi$.  (Note that
$\Ghat \xi = \Rhat^2 dt$, $\Ghat \xi \xi = \Rhat^2$, and $G = -e^{2t}\, \Go$.)
Generalizing, let $\Ghat$ now be {\em any} space-time metric of signature
$+---$ defined on a manifold $\hat{\cal M}$ on which there is a time-like
vector field $\xi$ with respect to which $\Ghat$ has the exponential expansion
property.  One can show that (locally, at least) there exist on $\hat{\cal M}$
coordinate systems ${[\![ t, x^m ]{\! ]}}$ for which
$\xi = \partial / \partial t$ and $\Ghat$ takes the form
\begin{mathletters}
\beqa
\Ghat &=& \phi^2 \bigl(dt + \Ao\bigr) \ox
                 \bigl(dt + \Ao\bigr) - e^{2t}\, \Go  \\
      &=& \phi^2 \bigl(dt + \Ao_m\, dx^m\bigr) \ox
                 \bigl(dt + \Ao_n\, dx^n\bigr) \nonumber \\
      & & {} - e^{2t} \bigl(dx^m \ox \go_{mn}\, dx^n\bigr),
\eeqa
\end{mathletters}
with $\phi$, $\Ao_m$, and $\go_{mn}$ independent of $t$.

The Ricci tensor $\Phihat$ and curvature scalar $\Psihat$ of $\Ghat$ are
expressible in terms of those of $\Go$ and covariant derivatives of $\phi$
and $\Ao$ with respect to $\Go$.  One can define `residuals'
$\Phihat_\infty$ and $\Psihat_\infty$ of $\Phihat$ and $\Psihat$ (roughly,
$\Phihat_\infty := \lim_{t \to \infty} (e^{-2t} \Phihat)$, and, exactly,
$\Psihat_\infty := \lim_{t \to \infty} \Psihat$).  One then finds that
$\Phihat_\infty = -3\, \phi^{-2} \Ghat$ and $\Psihat_\infty = -12\, \phi^{-2}$,
thus that
\beq
\Phihat_\infty - \case{1}{2} \Psihat_\infty \Ghat = \Lambda \Ghat,
\eeq
where $\Lambda := 3/\phi^2$.  Comparison with the de Sitter model shows that
the scalar field $\Lambda$ could be termed the `residual cosmological
(non)constant', and the scalar field $\phi$ the `residual (nonuniform) radius
of curvature', of the generalized model.  In the de Sitter model
$\Phihat v v = -\Lambda \Ghat v v$, which vanishes if $v$ is a null vector,
and is {\em negative} if $v$ is timelike.  Here the same is true of
$\Phihat_\infty$.

Field equations are obtained from the strictly geometric action principle
$\delta {\cal A} = 0$, where
\begin{mathletters}
\beqa
{\cal A}(\phi, \Ao_m)
    &:=& \int_{\hat{\cal D}} \bigl(\Psihat - \Psihat_\infty\bigr)\, d\Vhat \\
\vspace{5pt} \nonumber \\
 &\,\,=& \int_{\cal D} \int_a^b \bigl(\Psihat - \Psihat_\infty\bigr)\, dt\, dV,
\eeqa
\end{mathletters}
the region $\hat{\cal D}$ having the cylindrical form $\hat{\cal D} = [a, b]
\times \cal D$, where $\cal D$ is a bounded region of a cross section of
$\hat{\cal M}$ transverse to $\xi$.  The variations of $\phi$ and $\Ao_m$ are
to vanish on $[a, b] \times \partial \cal D$.  The spatial metric $\Go$ is
treated as given {\em a priori} on $\cal D$, and extended to $\hat{\cal D}$ by
translations along $\xi$.  Variation of $\phi$ yields the equation
\beq
\Ao^k\!{}_{:k} - \Ao^k \Ao_k 
       = \case{3}{8} \phi^2 \Fo_k {}^l \Fo_l {}^k - \case{1}{2} \Psio;
\label{varphi}
\eeq
variation of $\Ao_m$ yields
\beq
\Fo^{mk}\!{}_{:k} + 3\, \phi^{-1} \Fo^{mk} \phi_{.k}
       = 2\, \phi^{-2} \bigl(\phi^{-1} \phio^{.m} + 2 \Ao^m\bigr).
\label{varAo_m}
\eeq
Here $(\;)_{.m} := \partial (\;)/ \partial x^m$, $F_{mn} :=
\Ao_{n.m} - \Ao_{m.n}$, and insertion of a $\; \degree \;$ indicates raising of
an index by $\go^{mn}$.  The covariant differentiations indicated by a ${}_:$
are with respect to $\Go$.  A constant factor $e^{-(a + b)/2}$ arising from the
$t$ integration has been absorbed into $\phi$; this leaves in the equations
{\em no} arbitrary coupling constant with which to finesse the `energy
condition' question.

Examining these field equations for a metric of the spherically symmetric form
\beqa
\Ghat &=& e^{2 U(\rho)} \bigl[dt + V(\rho)\, d\rho\bigr]^2 \nonumber \\
      & & {} - e^{2t} e^{-3 U(\rho)}
                 \left[d\rho^2 + r^2(\rho)\, d\Omega^2\right],
\label{Ghat}
\eeqa
one finds them to be satisfied if
\begin{mathletters}
\label{ivproblem}
\beqa
&U'\; & = -2 V = {m \Rhat \over r^2}, \quad
    r'' = {1 - {r'}^2 \over {2 r}} - \frac78 {m^2 \Rhat^2 \over r^3}, \\
\vspace{15pt} \nonumber \\
&U&\!(\infty) = \ln \Rhat, \quad r(0) = r_0, \quad \text{and} \quad r'(0) = 0,
\eeqa
\end{mathletters}
where each of $m$, $\Rhat$, and $r_0$ is a constant, $\Rhat > 0$, and
$0 \leq m < m_{\text{crit}} := \bigl(2/\sqrt7\,\bigr)\bigl(r_0/\Rhat\bigr)$.

The coordinate changes $T := \Rhat\, \left[t + \int V(\rho)\, d\rho\right] =
\Rhat\, \left[t - \frac12 U(\rho)\right]$ and $\rhobar := \rho/\Rhat$ make
\beq
\Ghat = e^{2 \Ubar(\rhobar)} dT^2
          - e^{2T/\Rhat} \bigl[e^{-2 \Ubar(\rhobar)} d\rhobar^2
                                 + \rbar^2(\rhobar)\, d\Omega^2\bigr],
\eeq
where $\Ubar(\rhobar) := U(\rho) - \ln \Rhat$ and
$\rbar(\rhobar) := e^{-U(\rho)} r(\rho)$.

On a human time scale the cosmological expansion factor $e^{2T/\Rhat}$ can be
treated as a constant, say $e^{2T_0/\Rhat}$, and absorbed into the spatial
metric by the transformations $\rhotil := e^{T_0/\Rhat} \rhobar$ and
$\rtil(\rhotil) := e^{T_0/\Rhat}\, \rbar(\rhobar)$, to produce
\vskip -2pt
\begin{mathletters}
\label{flowform2}
\beqa
\Ghat \approx \Ghat_{T_0}
                      &:=& e^{2 \Util(\rhotil)} dT^2
                             - e^{-2 \Util(\rhotil)} d\rhotil^2
                             - \rtil^2(\rhotil)\, d\Omega^2  \\
\rule{0pt}{11pt}
                   &\,\,=& \left[1 - u^2(\rhotil)\right]\, dT^2 \nonumber \\
                       & & {} - \left[1 - u^2(\rhotil)\right]^{-1} d\rhotil^2
                              - \rtil^2(\rhotil)\, d\Omega^2  \\
\rule{0pt}{14pt}
                   &\,\,=& d\ttil^{\,2} - \left[d\rhotil
                               - u(\rhotil)\, d\ttil\,\right]^2
                               - \rtil^2(\rhotil)\, d\Omega^2,
\label{Ghatapprox}
\eeqa
\end{mathletters}
\vskip -2pt
\noindent where $\Util(\rhotil) := \Ubar(\rhobar)$,
$u(\rhotil) := -\sqrt{1 - e^{2 \Util(\rhotil)}}$, and
$\ttil := T - \int u(\rhotil) \bigl[1 - u^2(\rhotil)\bigr]^{-1}\, d\rhotil$.
Because Eqs.~(\ref{flowform2}b) and (\ref{flowform2}c) replicate
Eqs.~(\ref{schwarzform}) and (\ref{flowform1}), the previous discussion of
horizons, blackness, and singularities applies directly to the metric
$\Ghat_{T_0}$.

\begin{figure}[h]
\psfig{figure=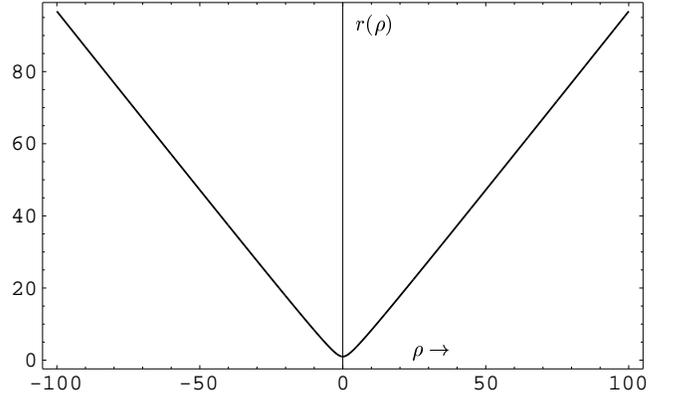,width=3.375truein}
\vskip 5pt

\caption{Plot of the spatial geometric descriptor $r(\rho)$
on the interval $-100 < \rho < 100$.}
\label{fig2}
\end{figure}

\begin{figure}[h]
\psfig{figure=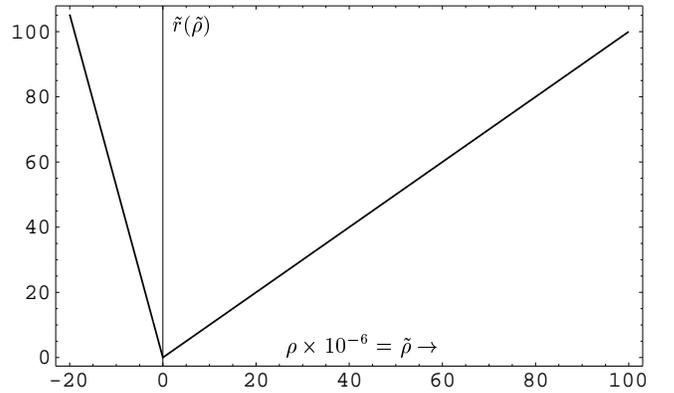,width=3.375truein}
\vskip 5pt

\caption{Plot of the spatial geometric descriptor $\rtil(\rhotil)$,
for $T_0 = 0$, on the interval $-2 \times 10^7 < \rho < 10^8$.}
\label{fig3}
\end{figure}

\begin{figure}[h]
\psfig{figure=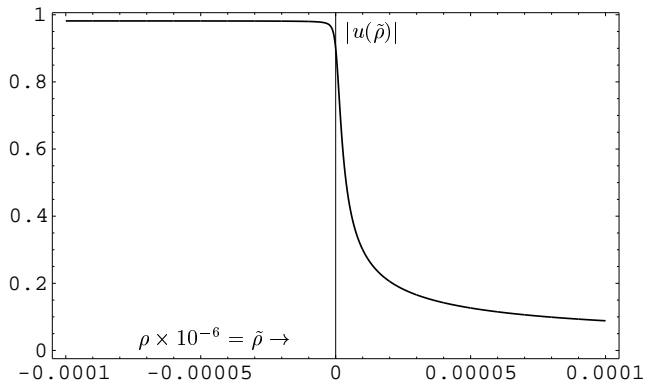,width=3.375truein}
\vskip 5pt

\caption{Plot of the test-particle free-fall speed $|u(\rhotil)|$, for
$T_0 = 0$, on the interval $-100 < \rho < 100$.}
\label{fig4}
\end{figure}

\begin{figure}[h]
\vskip -5pt
\psfig{figure=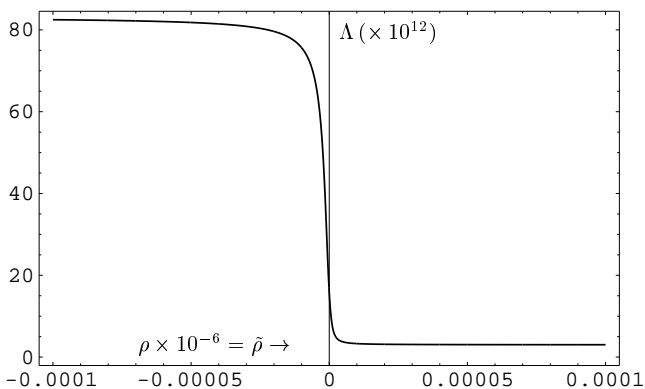,width=3.375truein}
\vskip 5pt

\caption{Plot of the `residual cosmological (non)constant' $\Lambda$ versus
$\rhotil$, for $T_0 = 0$, on the interval $-100 < \rho < 100$
($\Lambda := 3/\phi^2 = 3 e^{-2 U(\rho)}$).}
\label{fig5}
\end{figure}

Numerical integration produces the plots shown in
Figs.~(\ref{fig2}--\ref{fig5}), for which $r_0 = 1$, $\Rhat = 10^6$,
$m_{\text{crit}} \approx 7.6 \times 10^{-7}$, $m = 0.5\, m_{\text{crit}}$, and
$T_0 = 0$.  The minimum of $r = r(0) = r_0 = 1$, whereas
$\rtil_{\text{min}} \approx \rtil (9.3 \times 10^{-7}) \approx 1.93
\times 10^{-6}$.

Equations (\ref{ivproblem}) can also be integrated by hand (see Appendix).
The result is that
\beqa
\lefteqn
{
 \sgn(r'(\rho))\,\rho = \sqrt{(r - r_0)(r - \gamma^2 r_0)}
} \qquad \nonumber \\
  & & {} + (1 + \gamma^2) r_0 \ln \left({\sqrt{r - r_0}
                                          + \sqrt{r - \gamma^2 r_0}\;} \over
                                            \sqrt{{(1 - \gamma^2) r_0}}\right),
\label{rimplicit}
\eeqa
where $\gamma := m/m_{\text{crit}} < 1$, and
\beqa
\lefteqn
{
 U(\rho) = \ln \Rhat
} \qquad \quad \nonumber \\
  & & {} + \frac{4}{\sqrt7} \ln \left({\sqrt{r - \gamma^2 r_0}
                                   + \sgn(\rho) \gamma \sqrt{r - r_0}\;}
                                   \over {(1 + \gamma) \sqrt{r}}\right).
\label{Uequation}
\eeqa
Equation (\ref{rimplicit}) implicitly defines $r$ as a function of $\rho$ on
the interval $-\infty < \rho < \infty$, with minimum value $r(0) = r_0$, and
with $\sgn(r'(\rho)) = \sgn(\rho)$.  It is clear from this equation that, as
$\rho \to \pm \infty$, $r(\rho) \sim \infty$, $\rho/r(\rho) \sim \pm 1$, and,
consequently, $r(\rho) \sim \pm \rho$.  From this and $U' = m \Rhat/r^2$ it
follows that, as $\rho \to \infty$,

\begin{mathletters}
\beqa
U(\rho) &=& U(\infty)
             + \int_{\infty}^{\rho} \frac{m \Rhat}{r^2(\lambda)} \, d\lambda \\
     &\sim& U(\infty)
             + \int_{\infty}^{\rho} \frac{m \Rhat}{\lambda^2} \, d\lambda
           = \ln \Rhat - \frac{m \Rhat}{\rho},
\eeqa
\end{mathletters}
\noindent and, as $\rho \to -\infty$,
\vskip -3pt
\begin{mathletters}
\beqa
U(\rho) &=& U(-\infty)
            + \int_{-\infty}^{\rho} \frac{m \Rhat}{r^2(\lambda)} \, d\lambda \\
     &\sim& U(-\infty)
            + \int_{-\infty}^{\rho} \frac{m \Rhat}{\lambda^2} \, d\lambda 
                                                                   \nonumber \\
        & & = \ln \Rhat
              + \frac{4}{\sqrt7} \ln \left(\frac{1 - \gamma}
                                                {1 + \gamma}\right)
              - \frac{m \Rhat}{\rho}.
\eeqa
\end{mathletters}
Further,
\beqa
\lefteqn
{
 \frac{\rtil(\rhotil)}{\rhotil}
   = \frac{\rbar(\rhobar)}{\rhobar} = \Rhat e^{-U(\rho)} \frac{r(\rho)}{\rho}
} \qquad \qquad \nonumber \\
\nonumber \\
 &\sim& \left\{ \begin{array}{ll}
                  1 & \text{as $\rho \to \infty$,} \\
                 \displaystyle{
                  -\left(\frac{1 + \gamma}{1 - \gamma}\right)^{4/\sqrt7}}
                    & \text{as $\rho \to -\infty$.}
                \end{array}
        \right.
\label{rtilasymp}
\eeqa

\vskip -5pt

\noindent Also,
\vskip -5pt
\beqa
\lefteqn
{
 u^2(\rhotil) = 1 - e^{2 \Util(\rhotil)}
                      = 1 - e^{2 \Ubar(\rhobar)}
                      = 1 - e^{2 [U(\rho) - \ln \Rhat]}
} \nonumber \\
  \nonumber \\
 & & \sim \left\{ \begin{array}{ll}
                 \displaystyle{
                 \frac{2 m_{T_0}}{\rhotil}}
                   & \text{as $\rho \to \infty$,} \\
                 \displaystyle{
                 1 - \left(\frac{1 - \gamma}{1 + \gamma}\right)^{8/\sqrt7}\!
                     \left(1 - \frac{2 m_{T_0}}{\rhotil}\right)}
                   & \text{as $\rho \to -\infty$,}
                \end{array}
        \right.
\label{uasymp}
\eeqa
where $m_{T_0} := m e^{T_0/\Rhat}$, and
\beqa
\lefteqn
{
 \Lambda = 3 e^{-2 U(\rho)}
} \nonumber \\
  \nonumber \\
 & & \sim
        \left\{ \begin{array}{ll}
                 \displaystyle{
                 \frac{3}{\Rhat^2}\left(1 + \frac{2 m_{T_0}}{\rhotil}\right)}
                   & \text{as $\rho \to \infty$,} \\
                \\
                 \displaystyle{
                 \frac{3}{\Rhat^2} \left(\frac{1 + \gamma}
                                              {1 - \gamma}\right)^{8/\sqrt7}\!
                     \left(1 + \frac{2 m_{T_0}}{\rhotil}\right)}
                   & \text{as $\rho \to -\infty$.}
                \end{array}
        \right.
\eeqa
\vskip 5pt

On each time-slice of constant $\ttil$ the line element induced by
$\Ghat_{T_0}$ is $d\rhotil^2 + \rtil^2(\rhotil)\, d\Omega^2$
(see Eq.~(\ref{Ghatapprox})).  The fact that in the numerical solution
$\rtil(\rhotil)$ (consequently, also $\rbar(\rhobar)$) has a positive minimum
value and is asymptotically infinite at $\pm \infty$ tells that in the universe
described approximately by $\Ghat_{T_0}$ and exactly by $\Ghat$ there
\linebreak
\vskip -5pt

\noindent is an ever present Einstein--Rosen `tunnel' connecting two regions of
asymptotically Euclidean topology \cite{ein&ros35}.\footnote{Einstein and Rosen
spoke of a `bridge', but `tunnel' seems to describe the topology better.} Is
this a one-way tunnel, or can it accommodate two-way traffic?  The answer lies
in the behavior of $u$.  If $|u(\rhotil)| \geq 1$ somewhere, the traffic is
one-way only and there is a blackhole in the vicinity.  If $|u(\rhotil)| < 1$
everywhere, then traffic is two-way and there is no blackhole.  Figure
\ref{fig4} says, ``Two-way traffic, no blackhole.'' The existence of this
two-way tunnel with no blackhole is generic for $\gamma < 1$, inasmuch as
$\rtil(\rhotil)$ has a positive minimum value (see Appendix) and, in view of
(\ref{uasymp}) and the monotonicity of $U(\rho)$ ($U' = m \Rhat/{r^2} > 0$),
$u^2(\rhotil)$ rises monotonically with decreasing $\rho$ from
$u^2(\infty) = 0$ to
$u^2(-\infty) = 1 - [(1 - \gamma)/(1 + \gamma)]^{8/\sqrt7} < 1$.

An immediate consequence of (\ref{rtilasymp}) and (\ref{uasymp}) is that
$\Ghat_{T_0}$ is asymptotic to the Schwarzschild metric of (active) mass
parameter $m_{T_0}$ as $\rho \to \infty$.  Additionally, from
$\phi = e^{U(\rho)} \sim e^{U(\infty)} = \Rhat$ and
$\Lambda = 3/\phi^2 \sim 3/{\Rhat^2}$ as $\rho \to \infty$, we see
that, far from the center of gravitation in the positive $\rho$ direction,
the residual radius of curvature $\phi$ and the residual cosmological
(non)constant $\Lambda$ are asymptotic to the radius of curvature and the
cosmological constant of the de~Sitter universe.

The vector field $\partial_{\ttil} + u(\rhotil)\, \partial_{\rhotil}$ is
geodesic for $\Ghat_{T_0}$; it is the velocity field of a cloud of test
particles free-falling downward from rest at $\infty$.  The speed
$|u(\rhotil)|$ of such a free-falling particle increases monotonically with
decreasing $\rhotil$ right through the tunnel, out the other side, and on to
$-\infty$.  This entails that the particle, once past the narrowest part, the
`throat', of the tunnel, behaves as if pushed away from it --- that the
gravitating center is repulsive on the other, low side of the throat.
Moreover, the repulsion is stronger than the attraction, by a ratio of mass
parameters equal to $[(1 + \gamma)/(1 - \gamma)]^{4/\sqrt7}$, which ratio
increases to $\infty$ as $m \to m_{\text{crit}}$ (see Appendix).  Because,
however, $|u(-\infty)| < 1$, an observer free-falling from rest at $\infty$
never reaches light speed.  With a sufficient means of propulsion the observer
could, at any point, turn back and join a cohort of test particles free-falling
upward to rest at $\infty$, flowing with the geodesic velocity field
$[1 + u^2(\rhotil)] [1 - u^2(\rhotil)]^{-1}\, \partial_{\ttil} \,-\,
u(\rhotil)\, \partial_{\rhotil}$.  If the propulsion failed, he could at least
shine a light whose photons would eventually arrive at $\infty$, redshifted by
an amount that is the greater the closer $|u(-\infty)|$ is to 1, which, in
view of (\ref{uasymp}), is the closer that $\gamma$ is to 1, thus the closer
that $m$ is to $m_{\text{crit}}$.  A topological hole in space gravitating in
such a way is to an observer on the high side a `darkhole', as dark as you
like, but never black.  To an observer on the low side it would be a
`brighthole', emanating blue-shifted light that came through the tunnel from
the high side.\footnote{For the full cosmological metric $\Ghat$ there would
come into play the phenomenon, first seen in the de Sitter universe, that all
nontachyonic test particles come asymptotically to rest at a point in space
as $T \to \infty$ \cite{sch56}.}

In Newtonian terms, the cloud of test particles falling with velocity
$\partial_{\ttil} + u(\rhotil)\, \partial_{\rhotil}$ would have `specific
kinetic energy' (kinetic energy per unit of inertial mass) ${\cal KE} =
\frac12 u^2(\rhotil)$.  On the other hand $-\frac12 u^2(\rhotil)$ can be
identified as the Newtonian `specific gravitational potential' $\cal V$, in
the sense that $-\partial{\cal V}/\partial\rhotil = -m_{T_0}/\rtil^2(\rhotil)$,
which is, one can see, the acceleration of the test particles in the cloud.
It then is automatic that ${\cal KE + V} = 0$.

The shape-mirroring between the graphs of $|u(\rhotil)|$ and $\Lambda$ in
Figs.~\ref{fig4} and \ref{fig5} is not accidental, inasmuch as
$\Lambda = 3/{\Rhat^2 \big(1 - u^2(\rhotil)\big) =
3/\Rhat^2 \big(1 - 2 |\cal V|\big)}$.  This
relationship says that, in the space-time described by the metric $\Ghat$,
not only is the analog of the cosmological constant not constant, it is
determined by the specific gravitational potential $\cal V$, and is smallest
where $|\cal V|$ is smallest, largest where $|\cal V|$ is largest.

A point to notice is that there is no upper bound on the mass parameter $m$.
The inequality $m < m_{\text{crit}} := \big(2/\sqrt7\,\big)\big(r_0/\Rhat\big)$
merely correlates $m$ and the hole-sizing parameter $r_0$; $m$ can grow to any
size, but $r_0$ must grow along with it.  Moreover, no matter how small or how
large the positive mass $m$ (or, equivalently, the asymptotic mass parameter 
$m_{T_0}$), there can be darkholes of that strength that are wide, with a slow
flow, because $r_0 \gg \left(\sqrt7/2\right) m \Rhat$ (so that
$\gamma \approx 0$ and $|u(-\infty)| \ll 1$), and darkholes that are narrow,
with a fast flow, because $r_0 \approx \left(\sqrt7/2\right) m \Rhat$ (so that
$\gamma \approx 1$ and $|u(-\infty)| \approx 1$).  The actual size of the hole
is determined by the minimum value $\rtil_{\text{min}}$ of $\rtil$, the area of
its smallest spherical cross section being $4 \pi \rtil_{\text{min}}^2$.  This
minimum radius is proportional to $r_0$, the relation being such that
$\rtil_{\text{min}}$ rises monotonically from $e^{T_0/\Rhat}\, r_0/\Rhat$ to
$\left(2/\sqrt7\,\right)\!\left(1 + \sqrt7/2\right)^{1 + 2/\sqrt7}
e^{T_0/\Rhat}\, r_0/\Rhat \approx 3.32\, e^{T_0/\Rhat}\, r_0/\Rhat$ as
$m_{T_0}$ is increased from $0$ to $e^{T_0/\Rhat}\, m_{\text{crit}}$
(concurrently, the value of $\rhotil$ at which the minimum occurs increases
from 0 to $\left[2/\sqrt7 + \ln\left(4/\sqrt7\,\right)\right] e^{T_0/\Rhat}\,
r_0/\Rhat \approx 1.17\, e^{T_0/\Rhat}\, r_0/\Rhat$; see Appendix).
Thus the radius of the hole is of the order of magnitude of
$e^{T_0/\Rhat}\, r_0/\Rhat$ for all admissible values of its asymptotic mass
parameter $m_{T_0}$.  That $\rtil_{\text{min}}$ grows larger with $m_{T_0}$
one can `explain' as follows: to make room for the increasing gravitational
flux, the tunnel from the dark side to the bright side expands as if the space
it is made of possessed elasticity in the directions transversal to the flow.
As well, the bright side can be said to grow infinitely more `roomy' in
comparison to the dark side as $m_{T_0} \to e^{T_0/\Rhat}\, m_{\text{crit}}$,
a consequence of the asymptotic behavior of $\rtil(\rhotil)/\rhotil$ for
$\rho \to -\infty$ (and $\gamma \to 1$) displayed in~(\ref{rtilasymp}).

Further to be noticed is that although $m$ was restricted to nonnegative
values, this restriction is not dictated by the mathematics.  Nothing essential
is lost by retaining it, however, for every solution of Eqs.~(\ref{ivproblem})
with $m < 0$ has a companion solution with $m > 0$ such that their respective
human time-scale metrics $\Ghat_{T_0}$ are mirror images of one another under
an isometry that reverses the sense of $\rho$ and maps the asymptotic regions
$\rho \sim \pm \infty$ of one of the space-time manifolds to the opposite
asymptotic regions of the other (see Appendix).  The darkhole and the
brighthole make an indivisible, organic whole, not affected by any pretense
that its `mass' is negative.

Finally, consider the question of `negative energy' and `exotic matter'.  To
a high-side observer at a reasonable distance from the center the darkhole is
just a normal gravitational attractor, able to exhibit all of the {\em visible}
features of a blackhole.  To a low-side observer the brighthole is repulsive,
and thus popularly termed `exotic'.  Is the energy density therefore positive
on the dark side and negative on the bright side?  In a strict sense the
question is not meaningful, inasmuch as Einstein's assumed relationship between
energy and geometry has been explicitly disallowed here, with no substitute put
in its place.  If one admits, however, that the Einstein tensor $\Ehat$, which
is conserved (that is, has covariant divergence zero), represents in some
fashion a connection between energy and geometry, then examination of $\Ehat$
is in order.  That tensor decomposes naturally into three parts, as follows:
\beq
\Ehat = \Ehat_{\text{expansion}} + \Ehat_{\text{gravity}}
                                 + \Ehat_{\text{space}}.
\label{Esplit}
\eeq
In terms of the orthonormal coframe system $\{\omegahat^{\mu}\}$, proper to the
class of (generally noninertial) observers at rest in the coordinate system
${\lbb}\rho, \vartheta, \varphi{\rbb}$, defined by
\begin{mathletters}
\label{omegahats}
\beqa
\omegahat^0 &:=& e^{U(\rho)}\, dT/\Rhat, \\
\omegahat^1 &:=& e^{T/\Rhat} e^{-U(\rho)}\, d\rho, \\
\omegahat^2 &:=& e^{T/\Rhat} e^{-U(\rho)} r(\rho)\, d\vartheta, \\
\omegahat^3 &:=& e^{T/\Rhat} e^{-U(\rho)} r(\rho) (\sin \vartheta)\, d\varphi,
\eeqa
\end{mathletters}
such that
\beq
\Ghat = \omegahat^0 \ox \omegahat^0 - \omegahat^1 \ox \omegahat^1
         - \omegahat^2 \ox \omegahat^2 - \omegahat^3 \ox \omegahat^3,
\eeq
these parts are expressed by (see Appendix)
\FL
\beqa
\lefteqn
{
 \Ehat_{\text{expansion}}
} \nonumber \\
  &=& \Lambda \Ghat + e^{-T/\Rhat} \frac{2m \Rhat}{r^2(\rho)}
                               \left[\omegahat^0 \ox \omegahat^1 +
                                     \omegahat^1 \ox \omegahat^0\right],
\label{Eexpansion}
\eeqa
\FL
\beqa
\lefteqn
{
 \Ehat_{\text{gravity}}
} \nonumber \\
  &=& e^{-2 T/\Rhat} e^{2 U(\rho)} \frac{3 m^2 \Rhat^2}{r^4(\rho)} \nonumber \\
  & & \quad \times \left[\omegahat^0 \ox \omegahat^0 +
                          \omegahat^1 \ox \omegahat^1 -
                          \omegahat^2 \ox \omegahat^2 -
                          \omegahat^3 \ox \omegahat^3\right],
\label{Egravity}
\eeqa
\FL
\beqa
\lefteqn
{
 \Ehat_{\text{space}}
} \nonumber \\
  &=& e^{-2 T/\Rhat} e^{2 U(\rho)}
      \frac{4 {r_0}^2 + 7 m^2 \Rhat^2}{8 r_0 r^3(\rho)} \nonumber \\
  & & \qquad \times \left[-2(\omegahat^1 \ox \omegahat^1) +
                              \omegahat^2 \ox \omegahat^2 +
                              \omegahat^3 \ox \omegahat^3\right].
\label{Espace}
\eeqa

Each of these parts is, one can show, {\em individually} conserved in the
same sense that their sum $\Ehat$ is conserved; thus each may be taken as
descriptive of a particular, separate aspect of the space-time.  The part
$\Ehat_{\text{expansion}}$ arises primarily from the exponential expansion
property of the metric, with some modification owed to the presence of the
gravitational \mbox{attractor--repeller}.  To the extent that energy can be
said to reside in that expansion, its density as it appears to the observers
at rest would presumably be the coefficient of $\omegahat^0 \ox \omegahat^0$
in $\Ehat_{\text{expansion}}$, which is $3 e^{-2 U(\rho)}$ ($= \Lambda$), a
positive quantity.  By way of comparison, the Einstein tensor $\Ehat_{T_0}$
of the metric $\Ghat_{T_0}$ has no counterpart to $\Ehat_{\text{expansion}}$;
it reduces to $\Ehat_{\text{gravity}} + \Ehat_{\text{space}}$, but with
$e^{T_0/\Rhat}$ in place of $e^{T/\Rhat}$, and $e^{-2 T_0/\Rhat}$ in place
of $e^{-2 T/\Rhat}$.

There is a clear separation of the primary sources of the energies, momenta,
stresses, strains, and pressures that the tensors $\Ehat_{\text{gravity}}$
and $\Ehat_{\text{space}}$ presumably display.  For $\Ehat_{\text{gravity}}$
that source is the gravity of the \mbox{attractor--repeller}:
$\Ehat_{\text{gravity}}$ is proportional to the square of the mass parameter
$m$ (and inversely proportional to $r^4(\rho)$).  For $\Ehat_{\text{space}}$
the primary source is the curvature of space: the three components of
$\Ehat_{\text{space}}$ are the parts of the sectional curvatures of the metric
$d\rho^2 + r^2(\rho)\, d\Omega^2$ that are inversely proportional to
$r^3(\rho)$, modified by the factors $e^{-2 T/\Rhat}$ and $e^{2 U(\rho)}$.

If the coefficient of $\omegahat^0 \ox \omegahat^0$ in $\Ehat_{\text{gravity}}$
is taken to be the energy density of the gravitational field of the
\mbox{attractor--repeller}, then that energy density is positive everywhere, on
the repulsive side as well as on the attractive side.  Moreover, it remains so
for all observers moving subluminally, there being no Lorentz boost from the
coframe system $\{\omegahat^{\mu}\}$ to a moving coframe system in which the
00 component of $\Ehat_{\text{gravity}}$ is not positive (a property shared by
the 00 component of $\Ehat_{\text{expansion}}$).

It is instructive to study the $m = 0$ case.  The metric reduces to
\beq
\Ghat = dT^2 - e^{2T/\Rhat}
               \bigl[d\rhobar^2 + \rbar^2(\rhobar)\, d\Omega^2\bigr],
\eeq
where now $\rhobar = \rho/\Rhat$ and $\rbar(\rhobar) = r(\rho)/\Rhat$.  There
is no center of attraction or repulsion, there is just the tunnel connecting
the two asymptotically Euclidean regions.  An observer can sit at rest wherever
and for so long as he pleases and experience as a (nominally) gravitational
effect only the ongoing cosmic expansion of the space around him.  The
Einstein tensor reduces to $\Ehat_{\text{expansion}} + \Ehat_{\text{space}}$,
with $\Ehat_{\text{expansion}} = (3/\Rhat^2) \Ghat$ and
\beqa
\lefteqn
{
 \Ehat_{\text{space}}
     =  e^{-2 T/\Rhat} \frac{\rbar_0}{2 \rbar^3(\rhobar)}
} \nonumber \\
  & & \qquad \qquad \quad \times \left[-2(\omegahat^1 \ox \omegahat^1) +
                                          \omegahat^2 \ox \omegahat^2 +
                                          \omegahat^3 \ox \omegahat^3\right],
\eeqa
where $\rbar_0 := r_0/\Rhat$.  The only nonzero energy density present is the
$3/\Rhat^2$ contributed by $\Ehat_{\text{expansion}}$.  An alternate way of
expressing $\Ehat_{\text{space}}$ is
\FL
\beqa
\lefteqn
{
 \Ehat_{\text{space}} = e^{-2 T/\Rhat}
  \left[-\kappa_{\vartheta \varphi}(\omegahat^1 \ox \omegahat^1)\right.
} \qquad \nonumber \\
  & & \qquad \qquad \quad \;\,
  \left.{} - \kappa_{\rho \varphi}(\omegahat^2 \ox \omegahat^2)
           - \kappa_{\rho \vartheta}(\omegahat^3 \ox \omegahat^3)\right],
\eeqa
where $\kappa_{\vartheta \varphi}$, $\kappa_{\rho \varphi}$, and
$\kappa_{\rho \vartheta}$ are the sectional curvatures of the spatial metric
$d\rhobar^2 + \rbar^2(\rhobar)\, d\Omega^2$ referred to the tangent subspaces
spanned by
$\{\partial_{\vartheta}, \partial_{ \varphi}\}$,
$\{\partial_{\rho}, \partial_{ \varphi}\}$, and
$\{\partial_{\rho}, \partial_{ \vartheta}\}$, respectively, given by
\beq
\kappa_{\vartheta \varphi}
  = \frac{1 - ({\rbar}'){}^2 (\rhobar)}{\rbar^2(\rhobar)}
  = \frac{\rbar_0}{\rbar^3 (\rhobar)}
\eeq
and
\beq
\kappa_{\rho \varphi} = \kappa_{\rho \vartheta}
                      = -\frac{\rbar''(\rhobar)}{\rbar(\rhobar)}
                      = -\frac{\rbar_0}{2 \rbar^3 (\rhobar)}.
\eeq
Thus the components of $\Ehat_{\text{space}}$ are just the curvatures of space
diluted by the factor $e^{-2 T/\Rhat}$ induced by the ongoing cosmic expansion.
If there is energy bound up in these components, it has nothing to do with any
gravity in the sense of attraction or repulsion, but only to do with stresses
and strains associated with the curvature of space ({\em not} space-time).  It
exists and contributes to the {\em inertial} mass of the tunnel, but it does
not gravitate, so it has no {\em active} gravitational mass equivalent.  Its
manifestation as inertial mass could be thought of as the resistance presented
by these stresses and strains to the deformations of space that would be
required if the tunnel were to move.  That two of these stresses and strains
are associated with negative sectional curvatures should cause no alarm,
especially in light of the fact that the field equations that produced $\Ghat$
are {\em vacuum} field equations, deriving as they do from the action principle
$\delta \int \bigl(\Psihat - \Psihat_\infty\bigr)\, d\Vhat = 0$, which is no
less geometrical in concept than the action principle
$\delta \int \Psi \, dV = 0$ that yields the Einstein vacuum field equations.
To hold that such curvatures are rare and are to be found only in exotic
circumstances, to hold, in other words, that Nature abhors a negatively
curved vacuum, is to presume to know more about Nature than Nature knows about
itself.

Taken together, these considerations suggest that some energy can be 
associated with gravity and some cannot, thus that not all energy `produces'
gravity (a consequence of which might be that the `cosmological constant
problem' \cite{weinberg89} does not exist).  Do they support in any way the
widely held belief that there are `exotic' relationships between energy and
geometry that justify calling the energy `negative'?  No! They do not.  Their
lesson is clear: Energy relates to geometry as it will ---
not as some uninvited adjectives say it must.

{\bf Note}.  The metric $\Ghat_{T_0}$ and the space-time it describes are in
all qualitative aspects identical with those derived and extensively analyzed
under Case III in my 1973 paper Ether flow through a drainhole: a particle
model in general relativity \cite{ellis73}.\footnote{A comparison of
Ref.~\cite{ellis73} with the present paper should take into account that the
Ricci and Einstein tensors of \cite{ellis73} are the negatives of those used
here.} What I there called a `drainhole' I would in the present context call a
darkhole, or, more accurately, a `darkhole--brighthole'.  The flowless (that
is, the massless, nongravitating) drainhole, whose metric has the form
$dt^2 - \bigl[d\rho^2 + \bigl(\rho^2 + n^2\bigr)\, d\Omega^2\bigr]$, was
later reinvented and put on exhibit by Morris and Thorne as an example of a
`traversible wormhole' \cite{mor&thor88,clement89}.

\appendix

\section*{}

The definitional conventions used for the curvature scalar $\Psi$, Ricci tensor
$\Phi$, and Riemann tensor $\Theta$ of a metric $G$ are the following, in which
$\omega^{\mu} = dx^{\mu}$, $e_{\mu} = \partial/\partial x^{\mu}$, and
$(\;)_{.\mu} := e_{\mu} (\;) = \partial (\;)/ \partial x^{\mu}$:
\beqa
     G &=& \omega^{\kappa} \ox g_{\kappa \lambda}\, \omega^{\lambda}, \\
G^{-1} &=& e_{\kappa} \ox g^{\kappa \lambda}\, e_{\lambda}, \\
  \Psi &=& \Phi_{\kappa}{}^{\kappa}
           = \Phi_{\kappa \lambda} g^{\lambda \kappa}, \\
  \Phi &=& \omega^{\kappa} \ox \Phi_{\kappa \lambda}\, \omega^{\lambda}
           = \omega^{\kappa} \ox \Theta_{\kappa}{}^{\mu}{}_{\lambda \mu}\,
                                                           \omega^{\lambda}, \\
\Theta &=& \omega^{\kappa} \ox 2 \left(d_{\wedge} \omega_{\kappa}{}^{\mu}
                                             - \omega_{\kappa}{}^{\pi} \wedge
                                               \omega_{\pi}{}^{\mu}\right)
                                                             \ox e_{\mu} \\
       &=& \omega^{\kappa} \ox
           \Theta_{\kappa}{}^{\mu}{}_{\lambda \nu}\, \omega^{\nu} \ox
           \omega^{\lambda} \ox e_{\mu}, \\
\Theta_{\kappa}{}^{\mu}{}_{\lambda \nu}
       &=& \Gamma_{\kappa}{}^{\mu}{}_{\nu.\lambda} -
           \Gamma_{\kappa}{}^{\mu}{}_{\lambda.\nu} \nonumber \\
       & & {} +
           \Gamma_{\kappa}{}^{\pi}{}_{\nu} \Gamma_{\pi}{}^{\mu}{}_{\lambda} -
           \Gamma_{\kappa}{}^{\pi}{}_{\lambda} \Gamma_{\pi}{}^{\mu}{}_{\nu},
\eeqa
where, with $\bold d$ denoting the torsionless covariant differentiation
consistent with $G$, the connection 1-forms $\omega_{\kappa}{}^{\mu}$ and
connection coefficients $\Gamma_{\kappa}{}^{\mu}{}_{\lambda}$ are determined
by ${\bold d} e_{\kappa} = \omega_{\kappa}{}^{\mu} \ox e_{\mu} =
\Gamma_{\kappa}{}^{\mu}{}_{\lambda}\, \omega^{\lambda} \ox e_{\mu} =
\{_{\kappa}{}^{\mu}{}_{\lambda}\}\, \omega^{\lambda} \ox e_{\mu} =
\frac12 (g_{\nu \lambda.\kappa} + g_{\kappa \nu.\lambda} -
g_{\kappa \lambda.\nu}) g^{\nu \mu}\, \omega^{\lambda} \ox e_{\mu}$.

\centerline{\rule{1.0in}{.1mm}}
\vskip 5pt

The second of Eqs.~(\ref{ivproblem}a) implies that 
$\bigl[r \bigl({r'}^2\bigr) - r - 7 m^2 \Rhat^2/{4 r}\bigr]' = 0$, thus that
\beqa
{r'}^2 &=& 1 - \frac{c}{r} + \frac74 \frac{m^2 \Rhat^2}{r^2}
\label{rprime} \\
       &=& \frac{(r - r_0)(r - \gamma^2 r_0)}{r^2},
\label{rprimefactored}
\eeqa
where $c = \bigl(r_0^2 + 7 m^2 \Rhat^2/4\bigr)\big/r_0$, as determined by the
initial conditions $r(0) = r_0$ and $r'(0) = 0$, and
$\gamma := m/m_{\text{crit}} < 1$, with
$m_{\text{crit}} := \left(2/\sqrt7\,\right)(r_0/\Rhat)$.  (Solutions with
$\gamma \geq 1$ exist, but are not considered here.)  The latter of
Eqs.~(\ref{ivproblem}a) becomes
\beq
r'' = \frac{r_0}{2 r^3} \left[(1 + \gamma^2)(r - r_0) +
                              (1 - \gamma^2) r_0\right],
\eeq
from which follows that $r''(0) = (1 - \gamma^2)/{2 r_0} > 0$, thus that $r$
has the minimum value $r_0$ at 0, and that $\sgn(r'(\rho)) = \sgn(\rho)$.
Equation (\ref{rprimefactored}) implies that
$r r' = \sgn(r') \sqrt{(r - r_0)(r - \gamma^2 r_0)}$, which in turn
implies that
\beq
\int_0^{\rho} \frac{r}{\sqrt{(r - r_0)(r - \gamma^2 r_0)}} \, dr
    = \int_0^{\rho} \sgn(r'(\lambda)) \, d\lambda. 
\eeq
Computation of these integrals yields Eq.~(\ref{rimplicit}).

Determination of $U(\rho)$ proceeds as follows:
\beqa
\lefteqn
{
 U(\rho) - U(\infty) = \int_{\infty}^{\rho} U'(\lambda) \, d\lambda
            = \int_{\infty}^{\rho} \frac{m \Rhat}{r^2(\lambda)} \, d\lambda
} \nonumber \\
  &=& m \Rhat \int_{\infty}^{\rho}
          \frac{\sgn(r')}{r \sqrt{(r - r_0)(r - \gamma^2 r_0)}} \, dr \\
  \nonumber \\
  &=& m \Rhat \left\{
           \begin{array}{ll}
            {\displaystyle
             \int_{\infty}^{\rho}
                 \frac{1}{r \sqrt{(r - r_0)(r - \gamma^2 r_0)}} \, dr}
               & \text{if $\rho \geq 0$,} \\
            \\
            {\displaystyle
             \int_{\infty}^0
                 \frac{1}{r \sqrt{(r - r_0)(r - \gamma^2 r_0)}} \, dr} \\
            \\
            {\displaystyle
              - \int_{0}^{\rho}
                 \frac{1}{r \sqrt{(r - r_0)(r - \gamma^2 r_0)}} \, dr}
               & \text{if $\rho \leq 0$,}
           \end{array} \!\!\!\!\!
              \right. \\
  \nonumber \\
  &=& \frac{2 m \Rhat}{\gamma \, r_0} \left\{
           \begin{array}{lr}
            {\displaystyle
             - \ln \left(\frac{\sqrt{r(\rho) - \gamma^2 r_0}
                                - \gamma \sqrt{r(p) - r_0}\,}
                              {(1 - \gamma) \sqrt{r(\rho)}}\right)} \\
               \qquad \qquad  \text{if $\rho \geq 0$,} \\
            \\
            {\displaystyle
             \hspace{10pt} \ln \left(\frac{\sqrt{r(\rho) - \gamma^2 r_0}
                                - \gamma \sqrt{r(p) - r_0}\,}
                              {(1 + \gamma) \sqrt{r(\rho)}}\right)} \\
               \qquad \qquad  \text{if $\rho \leq 0$,} \\
           \end{array} \!\!\!\!\!
              \right. \\
  \nonumber \\
  &=& \frac{4}{\sqrt7}
           \ln \left(\frac{\sqrt{r(\rho) - \gamma^2 r_0}
                                 + \sgn(\rho) \gamma \sqrt{r(p) - r_0}\,}
                          {(1 + \gamma) \sqrt{r(\rho)}}\right)\!. \!\!\!\!\!
\label{U-U(infty)}
\eeqa
Upon replacement of $U(-\infty)$ by $\ln \Rhat$, Eq.~(\ref{Uequation}) follows.

\centerline{\rule{1.0in}{.1mm}}
\vskip 5pt

{}From $\rtil(\rhotil) := e^{T_0/\Rhat}\, \rbar(\rhobar) :=
e^{T_0/\Rhat} e^{-U(\rho)} r(\rho)$ and $\rhotil := e^{T_0/\Rhat} \rhobar :=
e^{T_0/\Rhat} \rho/\Rhat$ one sees that $\rtil'(\rhotil) = \rbar'(\rhobar) =
\Rhat\, e^{-U(\rho)} \left[r'(\rho) - r(\rho) U'(\rho)\right]$, thus that
$\rtil'(\rhotil^*) = 0$ if and only if $r'(\rho^*)/r(\rho^*) = U'(\rho^*)$,
where $\rho^* := \Rhat\, e^{-T_0/\Rhat} \rhotil^*$.  Because
$U' = m \Rhat/r^2$, this condition is equivalent to
$r^2(\rho^*) {r'}^2(\rho^*) = m^2 \Rhat^2 = 4 \gamma^2 r_0^2/7$, which in view
of Eq.~(\ref{rprimefactored}) is equivalent to
\beq
r^2(\rho^*) - (1 + \gamma^2) r_0 r(\rho^*) + \case{3}{7}\, \gamma^2 r_0^2 = 0.
\eeq
This, together with $r \geq r_0$, entails that
\beq
r(\rho^*) = \frac{1}{2}
              \left[(1 + \gamma^2)
                     + \sqrt{(1 - \gamma^2)^2
                              + \frac{16}{7}\, \gamma^2}\, \right] r_0.
\label{r(rhostar)}
\eeq
As $\gamma$ goes from 0 to 1, $r(\rho^*)$ increases steadily from $r_0$ to
$\left(1 + 2/\sqrt7\,\right) r_0$.

Equations (\ref{rimplicit}) and (\ref{r(rhostar)}) imply
\beqa
\rho^* &=& \left[\frac{2}{\sqrt7}\, \gamma
                 + \frac{1}{2} \left(1 + \gamma^2\right) \right. \nonumber \\
       & & \left. \quad \times \ln\left({{\sqrt{(1 - \gamma^2)^2 +
                                                 \frac{16}{7}\, \gamma^2} +
                                                 \frac{4}{\sqrt7}\, \gamma}\,
                                          \over {1 + \gamma^2}} \right)
           \right] r_0.
\eeqa
{}From this it follows that $\rhotil^*$ increases steadily from 0 to
$\left[2/\sqrt7 + \ln\bigl(4/\sqrt7\,\bigr)\right] e^{T_0/\Rhat}\, r_0/\Rhat$
as $\gamma$ goes from 0 to 1.

By combining $\rtil(\rhotil) = e^{T_0/\Rhat} e^{-U(\rho)} r(\rho)$ with
Eqs.~(\ref{U-U(infty)}) and (\ref{r(rhostar)}), one finds that the minimum
radius $\rtil(\rhotil^*)$ ($=: \rtil_{\text{min}})$ increases monotonically
from $e^{T_0/\Rhat}\, r_0/\Rhat$ to
$\bigl(2/\sqrt7\,\bigr) \bigl(1 + \sqrt7/2\bigr)^{1 + 2/\sqrt7}
e^{T_0/\Rhat}\, r_0/\Rhat$ as $\gamma$ goes from 0 to~1, thus as $m$ is
increased from 0 to $m_{\text{crit}}$, and $m_{T_0}$ from 0 to
$e^{T_0/\Rhat} m_{\text{crit}}$.

\centerline{\rule{1.0in}{.1mm}}
\vskip 5pt

Applying the coordinate change $T := \Rhat\, \left[t - \frac12 U(\rho)\right]$
alters Eq.~(\ref{Ghat}) to
\beq
\Ghat = \frac{e^{2 U(\rho)}}{\Rhat^2}\, dT^2
          - e^{2T/\Rhat}\, e^{-2 U(\rho)} \left[d\rho^2
                                 + r^2(\rho)\, d\Omega^2\right],
\eeq
\vskip -10pt
\noindent which makes
\beq
\Ghat_{T_0} = \frac{e^{2 U(\rho)}}{\Rhat^2}\, dT^2
                - e^{2T_0/\Rhat}\, e^{-2 U(\rho)} \left[d\rho^2
                                         + r^2(\rho)\, d\Omega^2\right].
\label{Ghat0}
\eeq
Consider now a second solution of Eqs.~(\ref{ivproblem}) in the form of a
metric $\Gcheck$, defined on the same manifold $\hat{\cal M}$ that $\Ghat$
is defined on, but with respect to its own coordinate system
${\lbb}\tcheck, \rhocheck, \vartheta, \varphi{\rbb}$, and having its own
parameters $\mcheck$ and $\rcheck_0$ ($\Rhat$, $\vartheta$, and $\varphi$ being
the same as for $\Ghat$), with $0 \geq \mcheck > \mcheck_{\text{crit}} :=
-\bigl(2/\sqrt7\,\bigr)\bigl(\rcheck_0/\Rhat\bigr)$.  Just as for $\Ghat$, the
coordinate change
$\Tcheck := \Rhat\, \left[\tcheck - \frac12 \Ucheck(\rhocheck)\right]$
and the condition $\Ucheck' = -2 \Vcheck$ make
\beq
\Gcheck = \frac{e^{2 \Ucheck(\rhocheck)}}{\Rhat^2}\, d\Tcheck^2
          - e^{2\Tcheck/\Rhat}\, e^{-2 \Ucheck(\rhocheck)} \left[d\rhocheck^2
                                 + \rcheck^2(\rhocheck)\, d\Omega^2\right]
\eeq
\vskip -10pt
\noindent and
\beq
\Gcheck_{\Tcheck_0} = \frac{e^{2 \Ucheck(\rhocheck)}}{\Rhat^2}\, d\Tcheck^2
                       - e^{2\Tcheck_0/\Rhat}\,
                         e^{-2 \Ucheck(\rhocheck)} \left[d\rhocheck^2
                                     + \rcheck^2(\rhocheck)\, d\Omega^2\right].
\label{Gcheck0}
\eeq
Next suppose that $\Tcheck$ and $T$ are related by
\FL
\beqa
  \Tcheck &=& T \left(\frac{1 - \gamma}{1 + \gamma}\right)^{4/\sqrt7}, \\
\text{and $\rhocheck$ and $\rho$, by \qquad}
          & & \nonumber \\
\rhocheck &=& -\rho \left(\frac{1 + \gamma} {1 - \gamma}\right)^{4/\sqrt7},
\label{rhoreverse}
\eeqa
and let $\cal F$ be the diffeomorphism of $\hat{\cal M}$ that maps the point
$P$ with coordinates ${\lbb}T, \rho, \vartheta, \varphi{\rbb}$ to the point
${\cal F}(P)$ with coordinates
${\lbb}\Tcheck, \rhocheck, \vartheta, \varphi{\rbb}$, that is,
${\cal F} = \Xcheck^{-1} X$, where $X\!\!: \hat{\cal M} \to {\Bbb R}^4$ is
the coordinate system ${\lbb}T, \rho, \vartheta, \varphi{\rbb}$ and
$\Xcheck\!: \hat{\cal M} \to {\Bbb R}^4$ is the coordinate system
${\lbb}\Tcheck, \rhocheck, \vartheta, \varphi{\rbb}$.  For $\cal F$ to be an
isometry with respect to $\Ghat_{T_0}$ and $\Gcheck_{\Tcheck_0}$ it is
necessary and sufficient that the pullback of $\Gcheck_{\Tcheck_0}$ by $\cal F$
be equal to $\Ghat_{T_0}$, that is,
$\Gcheck_{\Tcheck_0}({\cal F})(d{\cal F})(d{\cal F}) = \Ghat_{T_0}$.
This will be true if and only if the expression
\beqa
\lefteqn
{
 \frac{e^{2 \Ucheck(\rhocheck)}}{\Rhat^2}
   \left(\frac{1 - \gamma}{1 + \gamma}\right)^{8/\sqrt7} dT^2
} \nonumber \\
  & & {} - e^{2\Tcheck_0/\Rhat}\, e^{-2 \Ucheck(\rhocheck)}
           \left[\left(\frac{1 + \gamma}
                            {1 - \gamma}\right)^{8/\sqrt7}\! d\rho^2
                                    + \rcheck^2(\rhocheck)\, d\Omega^2\right]
\eeqa
for $\Gcheck_{\Tcheck_0}$ derived from Eq.~(\ref{Gcheck0}) agrees with the
expression of $\Ghat_{T_0}$ in Eq.~(\ref{Ghat0}).  This in turn will be true
if and only if $\Tcheck_0 = T_0$,
\FL
\beqa
\Ucheck(\rhocheck) &:=& U(\rho) +
                   \ln \left(\frac{1 + \gamma}
                                  {1 - \gamma}\right)^{4/\sqrt7}, \\
\text{and \qquad \qquad} && \nonumber \\
\rcheck(\rhocheck) &:=& \left(\frac{1 + \gamma}
                                   {1 - \gamma}\right)^{4/\sqrt7} r(\rho).
\eeqa
(Note that $\rcheck(\rhocheck)/\rhocheck = -r(\rho)/\rho \to -1$ as
$\rho \to \infty$, thus as $\rhocheck \to -\infty$.) Are these consistent with
the supposition that $\Ucheck$ and $\rcheck$ satisfy Eqs.~(\ref{ivproblem})?
One has that
\beq
\Ucheck'(\rhocheck) = \frac{d\rho}{d\rhocheck} U'(\rho)
                    = \frac{d\rho}{d\rhocheck} \frac{m \Rhat}{r^2 (\rho)}
                    = -\left(\frac{1 + \gamma}
                                  {1 - \gamma}\right)^{4/\sqrt7}
                       \frac{m \Rhat}{\rcheck^2 (\rhocheck)},
\eeq
thus that $\Ucheck'(\rhocheck) = (\mcheck \Rhat)/\rcheck^2(\rhocheck)$
provided only that
\beq
\mcheck = -m \left(\frac{1 + \gamma}{1 - \gamma}\right)^{4/\sqrt7}.
\eeq
It is straightforward to check that this same condition guarantees that
$\rcheck$ will satisfy the second of Eqs.~(\ref{ivproblem}a).  Satisfaction of
Eqs.~(\ref{ivproblem}b) demands only the further stipulation that
$\rcheck_0 = r_0 \left[(1 + \gamma)/(1 - \gamma)\right]^{4/\sqrt7}$.

{}From these calculations the following inferences may be drawn:
\begin{enumerate}
\item For every space-time metric $\Ghat$ of the assumed form
(Eq.~(\ref{Ghat})) that satisfies the initial-value problem of
Eqs.~(\ref{ivproblem}) with a positive mass parameter $m$ there is one with a
negative mass parameter $\mcheck$ whose human time-scale approximant space-time 
metric $\Gcheck_{T_0}$ is, at each era $T_0$ of cosmic time, isometric to
$\Ghat_{T_0}$ --- and vice versa.  Consequently, there is on the human
time-scale no useful distinction to be made between the metrics with $m > 0$
and those with $m < 0$.
\item Each human time-scale approximant $\Ghat_{T_0}$ of the metric $\Ghat$ is
self-isometric under an isometry that reverses the direction of increase of
$\rho$ (cf. Eq.~(\ref{rhoreverse})), therefore maps the asymptotic region
$\rho \sim \infty$ onto the opposite asymptotic region $\rho \sim -\infty$.
Moreover, the details of that isometry make clear that, whereas $\Ghat_{T_0}$
is asymptotic as $\rho \to \infty$ to a Schwarzschild metric with positive mass
parameter $m_{T_0}$ $\bigl(:= m e^{T_0/\Rhat}\bigr)$, it is asymptotic as
$\rho \to -\infty$ to a Schwarzschild metric with negative mass parameter
$\mcheck_{T_0}$ $\bigl(:= \mcheck e^{T_0/\Rhat}\bigr)$ such that
\beq
\frac{-\mcheck_{T_0}}{m_{T_0}} = \frac{-\mcheck}{m}
                               = \left(\frac{1 + \gamma}
                                            {1 - \gamma}\right)^{4/\sqrt7} > 1.
\eeq
\end{enumerate}

\centerline{\rule{1.0in}{.1mm}}
\vskip 5pt

Introduction of the coordinate $\That := t + \int V(\rho)\, d\rho$ gives the
metric of Eq.~(\ref{Ghat}) the (tensor product) form
\beqa
\lefteqn
{
 \Ghat = e^{2 U(\rho)} (d\That \ox d\That\,)
          - e^{2 \That} e^{-2 \int V(\rho)\, d\rho} e^{-3 U(\rho)}
} \nonumber \\
  & & \qquad \qquad \times \left(\omega^1 \ox \omega^1
                               + \omega^2 \ox \omega^2
                               + \omega^3 \ox \omega^3\right),
\eeqa
where $\omega^1 := d\rho$, $\omega^2 := r(\rho) d\vartheta$, and
$\omega^3 := r(\rho) (\sin \vartheta) d\varphi$.  A standard calculation of the
Einstein tensor $\Ehat$, followed by application of Eqs.~(\ref{ivproblem}a),
yields the following equation, provided that $C = 0$ is chosen when the
antidifferentiation $\int V(\rho)\, d\rho = -\frac12 \int U'(\rho)\, d\rho =
-\frac12 U(\rho) + C$ is performed:
\beqa
\Ehat &:=& \Phihat - \case12 \Psihat \Ghat \nonumber \\
   &\,\,=& 3 e^{-2 U(\rho)} \Ghat
            + e^{-2 \That} e^{4 U(\rho)}
              \left[\frac34 \frac{m^2 \Rhat^2}{r^4(\rho)}\right]
              \left(d\That \ox d\That\right) \nonumber \\
       & & {} + \frac{2m \Rhat}{r^2(\rho)} \left(d\That \ox \omega^1 +
                                       \omega^1 \ox d\That\right) \nonumber \\
       & & {} + \left[-\frac{1 - {r'}^2(\rho)}{r^2(\rho)}
                       - \frac{m^2 \Rhat^2}{r^4(\rho)}\right]
                  \left(\omega^1 \ox \omega^1\right) \nonumber \\
       & & {} + \left[\frac{1 - {r'}^2(\rho)}{2 r^2(\rho)}
                       + \frac18  \frac{m^2 \Rhat^2}{r^4(\rho)}\right]
                  \left(\omega^2 \ox \omega^2 +
                        \omega^3 \ox \omega^3\right). \nonumber \\
       & &
\eeqa
Using the definitions of Eqs.~(\ref{omegahats}) in this equation, as well as
Eq.~(\ref{rprime}), one arrives at the decomposition of $\Ehat$ expressed in
Eqs.~(\ref{Esplit}) and (\ref{Eexpansion}--\ref{Espace}).

\vskip 30pt
\noindent Homer G. Ellis \\
Department of Mathematics \\
University of Colorado at Boulder \\
Boulder, CO 80309-0395 \\

\noindent Email: ellis@euclid.colorado.edu \\
Telephone: (303) 492-7754 (office) \\
\phantom{Telephone: }(303) 499-4027 (home) \\
Fax: (303) 492-7707

\end{document}